\documentclass[rnote]{aa}
\usepackage{txfonts}
\usepackage{graphicx}
\usepackage{aalongtable}
\begin{document}

\title{Interferometric observations of $\eta$ Carinae with VINCI/VLTI}

\author{P. Kervella}

\institute{LESIA, UMR 8109, Observatoire de Paris-Meudon, 5, place Jules Janssen, 
F-92195 Meudon Cedex, France}

\titlerunning{Interferometric observations of $\eta$ Car with VINCI/VLTI}
\authorrunning{P. Kervella}
\mail{pierre.kervella@obspm.fr}

\date{Received ; Accepted }

\abstract
{The bright star $\eta$ Carinae is the most massive and luminous star in our region of
the Milky Way. Though it has been extensively studied using many different techniques,
its physical nature and the mechanism that led to the creation of the Homunculus nebula
are still debated.}
{We aimed at resolving the central engine of the $\eta$~Carinae complex in the near-infrared
on angular scales of a few milliarcseconds.}
{We used the VINCI instrument of the VLTI to recombine coherently the light from
two telescopes in the $K$ band.}
{We report a total of 142 visibility measurements of $\eta$~Car, part of which
were analyzed by Van Boekel et al.~(2003).
These observations were carried out on projected baselines ranging from 8 to 112 meters
in length, using either two 0.35\,m siderostats or two 8-meter Unit Telescopes.
These observations cover the November 2001 - January 2004 period.}
{The reported visibility data are in satisfactory agreement with the recent results obtained with
AMBER/VLTI by Weigelt et al.~(2006), asuming that the flux of $\eta$~Car
encircled within 70\,mas reaches 56\% of the total flux within 1400\,mas, in the $K$ band.
We also confirm that the squared visibility curve of $\eta$~Car as a function of spatial frequency
follows closely an exponential model.}


\maketitle

\section{Introduction}
$\eta$ Carinae, the brightest example of the S Doradus class of stars, is the
most massive, most luminous star in our region of
the Milky Way. Over the last two hundred years $\eta$~Car has
shown many signs of violent activity, with in particular a spectacular eruption in the 1840s
that created the Homunculus nebula. The study of $\eta$~Car raises important
questions about how the most massive stars may end their lives. The central
object was studied by Weigelt \& Ebersberger~(\cite{weigelt86}) and
Falcke et al.~(\cite{falcke96}) using speckle interferometry at an angular resolution of
the order of 30\,milliarcsecond (mas). This revealed a complex structure with several equatorial
blobs at distances of 0.1 to 2 arcsec from the star, but the central engine remained
unresolved. Long baseline interferometry, currently the only technique allowing the
mas resolution necessary to resolve $\eta$~Car, was recently applied to
this star in the near- and mid-infrared domains by Van Boekel et al.~(\cite{vanboekel03}),
Chesneau et al.~(\cite{chesneau05}) and Weigelt et al.~(\cite{weigelt06}).
At the estimated distance of $\eta$~Car of 2.3\,kpc (Davidson \& Humphreys~\cite{davidson97},
Davidson et al.~\cite{davidson01}, Smith~\cite{smith06}), one mas corresponds to 2.3\,AU.
We report in this Research Note the complete corpus of VINCI observations of
$\eta$~Car in the $K$ band, including those discussed
by Van Boekel et al.~(\cite{vanboekel03}).

\section{Observations}\label{vis_values}
The Very Large Telescope Interferometer
(VLTI, Glindemann et al.~\cite{glindemann03b})
has been operated by the
European Southern Observatory on
top of the Cerro Paranal, in Northern Chile since March 2001.
For the present work, the light from $\eta$~Car and its calibrators was
collected either by two 0.35m VLTI Test Siderostats or two 8\,m Unit Telescopes (UTs)
without adaptive optics. It was subsequently recombined coherently in the VINCI
instrument using a $K$ band filter ($\lambda=2.0-2.4\,\mu$m).

We have observed $\eta$~Car repeatedly over the period November 2001 to January 2004.
This resulted in a total of 71\,000 interferograms on this target, out of which 50\% (35\,639)
were selected automatically by the pipeline.
Approximately the same quantity of data were obtained on the calibrators.
We used the standard VINCI data reduction pipeline
(Kervella et al.~\cite{kervella04a}, version 3.1) to derive instrumental visibilities.
The calibration of $\eta$~Car's visibilities was done using well-known reference
stars selected in the Bord\'e et al.~Ê(\cite{borde02}) and Cohen et al.~(\cite{cohen99})
catalogues, except $\beta$\,Car.
The diameter of $\beta$\,Car
was computed from an interferometric measurement obtained with the
Intensity Interferometer (Hanbury Brown et al~\cite{hanbury74}).
The original $V$ band uniform disk (UD)
angular diameter was converted into a $K$ band uniform
disk angular diameter ($\theta_{\rm UD} = 1.54 \pm 0.10$\,mas)
using linear limb darkening coefficients from Claret et al. (\cite{claret00}).
Thanks to the relatively low values of $\eta$~Car's
visibilities, the systematic uncertainty due to the calibrators is in general a
small fraction of the total error bars.

\begin{figure}[t]
\centering
\includegraphics[bb=0 0 360 360, width=8.7cm]{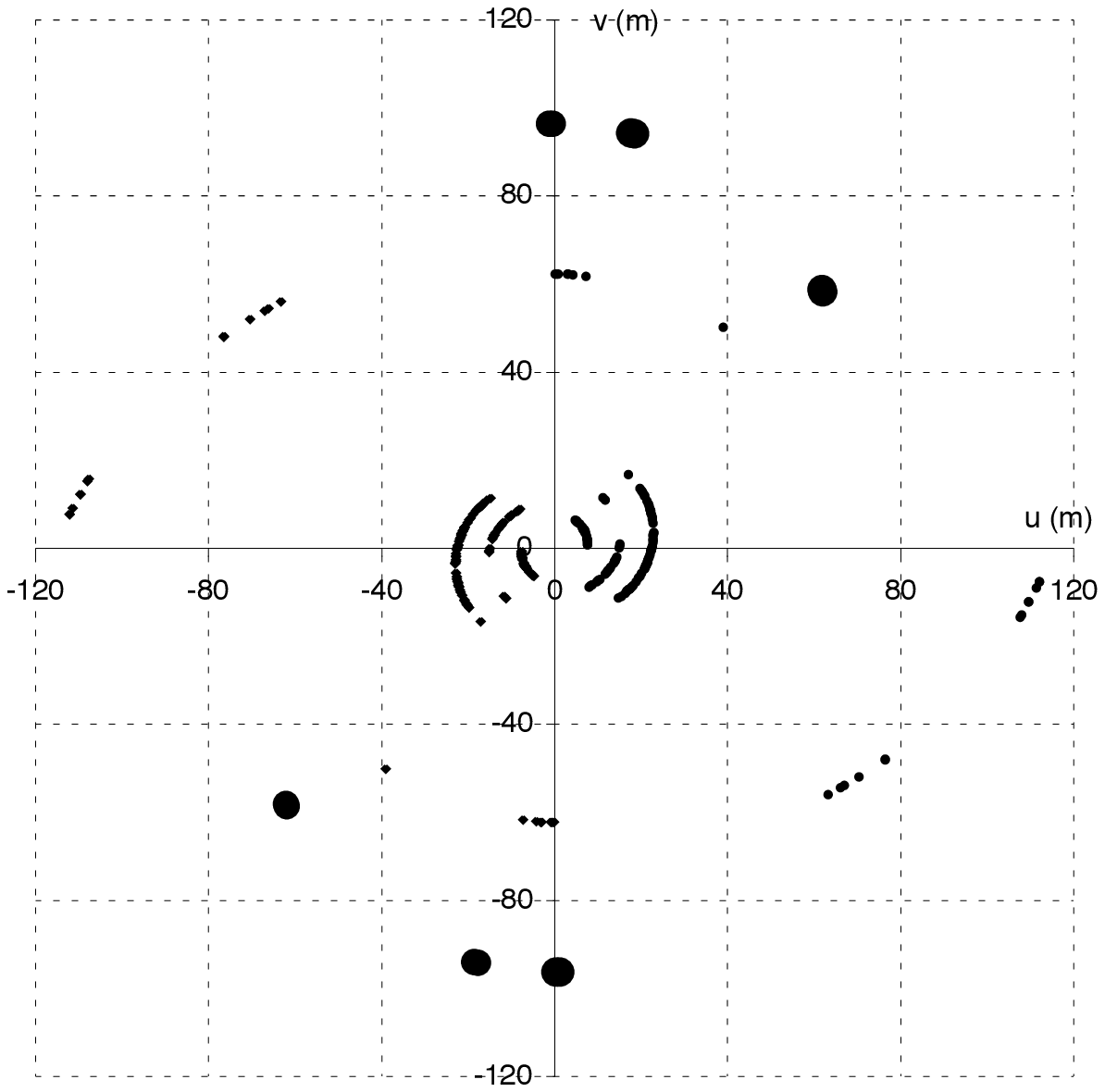}
\caption{$(u,v)$ coverage of the $\eta$~Car observations. Large spots represent
the UT observations and small spots the siderostat observations (North is up and East is to the right).}
\label{eta_car_uv}
\end{figure}

The calibrated visibility values obtained on $\eta$~Car are listed in Table~\ref{table_vis}.
Thanks to the use of several different telescope configurations
and to the supersynthesis effect, we were able to cover a broad range of 
baseline lengths and azimuth. The $(u,v)$ coverage of our observations is presented in
Figure~\ref{eta_car_uv}.
 
\section{Effective wavelength \label{wavelength}}
The VINCI instrument has no spectral resolution and its bandpass
corresponds to the $K$ band filter (2.0-2.4 $\mu$m). It is thus important to
compute the precise effective wavelength of the instrument in order to determine the 
spatial frequency of the observation. The true effective wavelength
differs from the filter mean wavelength mainly because of the object spectrum shape,
the detector quantum efficiency, and the fiber beam combiner transmission.
To derive the effective wavelength of our observations, we computed
a model taking into account $\eta$~Car's spectrum.
The instrumental transmission of VINCI and the VLTI was measured on bright reference stars
with the UTs (see Kervella et al. \cite{kervella03b} for details).
Due to the extraordinarily dense, opaque stellar wind, the shape
of the $\eta$~Car spectrum in the infrared is different from the
curve of a black body at the effective temperature of the central object.
In particular, the flux is increasing by about 20\% from 2.0 to 2.5~$\mu$m (Smith~\cite{smith02b}).
In our model, no spectral line either in emission or absorption has been taken into
account, considering the relatively limited contribution of these spectral features
to the total flux in the $K$ band.
Taking the average wavelength of this model spectrum gives an effective wavelength of
$\lambda_{\mathrm{eff}} = 2.196\ \mu$m for our $\eta$~Car observations, slightly longer than the typical
2.179 $\mu$m value for solar-type stars. We estimate the uncertainty on this effective
wavelength to less than $\pm 0.5$\%, or $\pm 0.01 \mu$m.

\section{Interferometric field of view}\label{fov_uts}
When injecting the light from an extended astronomical object into a single-mode fiber,
the wavefront corrugation by the atmosphere (loss of coherence) is converted into photometric
fluctuations. They are easily corrected during the data processing using the dedicated photometric
channels of VINCI. Unfortunately, the restoration of coherence by spatial filtering
comes at the expense of a very small field of view (FOV).
It is well approximated by the diffraction pattern of a telescope whose
size is the geometric mean of the apertures of the two telescopes.
In the case of our homogeneous two-UT observations, the FOV is thus
70 mas in the $K$ band.
Considering the extension of the $\eta$~Car complex, this limited FOV has an impact on
the measured visibilities.

Guyon (\cite{guyon}) studied in detail this limitation for the interferometric observation of extended
objects. One important conclusion is that the effective FOV depends on
the seeing, and so does the visibility. This is particularly true when large telescopes are
used without adaptive optics, as this was the case for our observations.
While all the UTs are now equipped with MACAO adaptive optics systems (Arsenault et al.~\cite{arsenault04}),
the early observations reported here were all obtained with atmosphere limited point spread functions.
The atmospheric turbulence creates a large cloud of speckles on the fiber head, and incoherent light
coming from separate parts of the object is coupled into the fiber, therefore reducing the
contrast of the fringes.
As a second order effect, different local seeing conditions for the two UTs could also slightly degrade
the visibilities.
In the case of small objects such as single stars, this effect is negligible, but $\eta$~Car is surrounded
by a large and bright envelope that is resolved by the UTs and contributes significantly
to the light distribution within the FOV.

Practically, this means that the visibility measurements obtained with the UTs
should be debiased from the seeing fluctuations. Unfortunately, this is not an easy task
because the relationship between the speckle cloud size (defined by the seeing)
and the flux coupled into the optical fiber is unknown. Tentatively, we
mention as a first estimation of the UTs FOV the observatory seeing in the $K$ band
at the time of the observations. The seeing values from the Paranal DIMM,
obtained at $\lambda = 0.5\ \mu$m have been converted to the $K$ band assuming
a classical $\lambda^{-6/5}$ dependance. Future comparisons of the visibility measurements
reported in the present Note with results from other instruments should take into account their
relative interferometric FOV. 

On the other hand, the observations obtained with the 0.35m
siderostats are in principle not affected by this
bias because most of the $\eta$~Car flux is coming from an area on the sky that
is contained into the Airy pattern of these telescopes.
Therefore, the obtained visibility is expected to be a faithful measurement of $\eta$~Car's
intrinsic visibility in the 1.40 arcsec FOV of the siderostats.
For the E0-G1 baseline, many visibility points have been obtained on different nights,
with a broad range of seeing conditions. The fact that they give very consistent
visibility values is a confirmation that the FOV variation
is negligible for the siderostats. 

\section{Discussion}

\begin{figure}[t]
\centering
\includegraphics[bb=0 0 360 288, width=8.7cm]{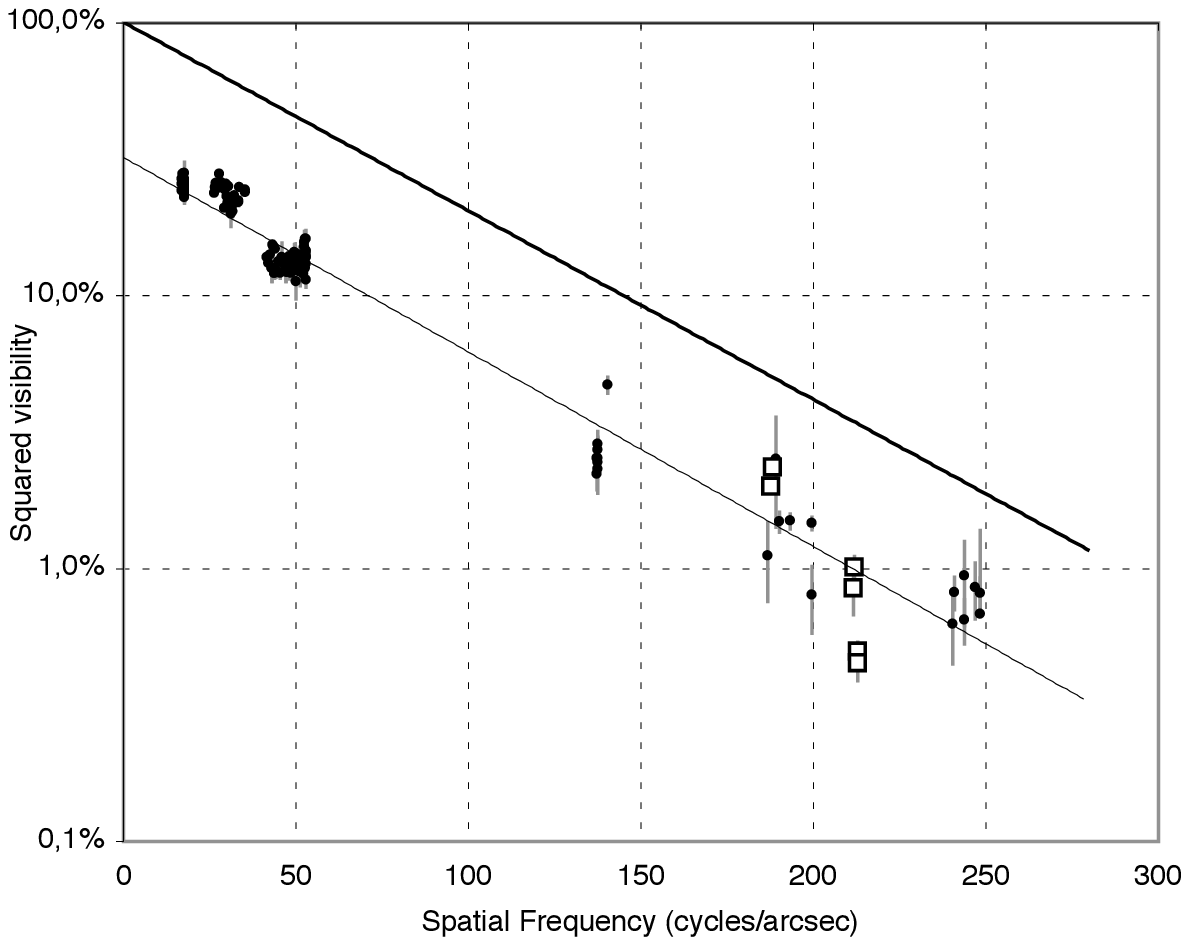}
\caption{Squared visibilities obtained on $\eta$~Car with VINCI, compared to the
model fitting result of Weigelt et al.~(\cite{weigelt06}), represented
as a solid curve. The UT data points are represented with open squares.}
\label{eta_car_vis}
\end{figure}

Figure~\ref{eta_car_vis} shows a comparison of the VINCI squared visibilities with the
AMBER model fitting result of Weigelt et al.~(\cite{weigelt06}), represented as a thick curve.
The VINCI squared visibilities show a strong decrease with increasing spatial frequencies,
clearly indicating that the central source is resolved by the interferometer.
The measurements obtained with the UTs, though in principle
affected by an uncertainty due to the variation of the FOV with the seeing,
are roughly consistent with the siderostat data obtained on comparable baselines.
The simple model developed by Hillier et al.~(\cite{hillier01}, \cite{hillier06}), was adjusted
by Weigelt et al.~(\cite{weigelt06}) to the AMBER observations of $\eta$~Car in the continuum
at $\lambda=2.174\,\mu$m. This model is well reproduced by an exponential curve
following the expression:
\begin{equation}
V^2 = 1.008\,\exp(-0.016\ s),
\end{equation}
where $s=B/\lambda$ is the spatial frequency.
Our wavelength reference is $\lambda=2.196\,\mu$m (Sect.~\ref{wavelength}).
On the same figure, the dashed curve is an exponential fit to the VINCI data:
\begin{equation}
V^2 = 0.322\,\exp(-0.016\ s).
\end{equation}
The slopes (in logarithmic scale)
of the VINCI data fit and the model representing the AMBER measurements are in excellent agreement.
However, the ratio of the two (VINCI/AMBER) is $\rho^2 = 32$\% in squared visibilities,
translating into a factor $\rho = 56$\% in visibilities. This ratio is
constant with the spatial frequency, the signature of a fully resolved component.

To estimate the contribution of this extended component, we can consider the
FOV of the two instruments.
While the AMBER observations were obtained with the MACAO adaptive optics system in function
(the FOV was thus $\approx$70\,mas), the FOV of the VINCI siderostat observations
was much larger, $\approx$1400\,mas.
From the observed ratio $\rho$ between the visibilities measured by VINCI and
AMBER, we can infer that 56\% of the 1400\,mas encircled $K$ band
flux of $\eta$~Car comes from within the 70\,mas point spread function of a single UT.
This value is nicely consistent with the independent measurement by
Van Boekel et al.~(\cite{vanboekel03}), based on adaptive optics observations with the
NACO instrument, that gives an encircled energy of 57\% within 70\,mas.
When corrected for the contribution of the extended emission, the visibilities
measured by AMBER and VINCI are in excellent agreement.

%
A discussion of the shape of the dense stellar wind of $\eta$\,Car
can be found in Smith et al.~(\cite{smith03}) and Van Boekel et al.~(\cite{vanboekel03}).
To improve the currently simplified spherical models,
this observable appears highly desirable.
The operating VLTI instruments are
now routinely providing spectro-interferometric datasets on $\eta$~Car
(Weigelt et al.~\cite{weigelt06}; Chesneau et al.~\cite{chesneau05}), and
the planned second generation will combine at least
four telescopes, allowing to obtain rich data cubes at mas scales.
This is an essential effort to follow the extremely fast evolution
of $\eta$~Car (Martin et al.~\cite{martin06}).
In this context, the simple, two-telescopes, broadband VINCI data
provide an interesting fiducial.

\begin{acknowledgements}
Based on observations made with ESO's VLT Interferometer at Cerro Paranal, Chile.
The VINCI data were retrieved from the ESO/ST-ECF Archive.
This research made use of the SIMBAD and VIZIER databases
at the CDS, Strasbourg (France), and of NASA's Astrophysics Data
System Bibliographic Services.
\end{acknowledgements}

{}


\begin{small}
\begin{longtable}{cccrrrrl}
\caption{Squared visibilities measured with VINCI on $\eta$~Car.
The seeing in the $K$ band at the time of the observation is given as the
FOV with the UTs (see Section \ref{fov_uts}). $N$ is the number of processsed interferograms,
$B$ the baseline length, and Az. the azimuth angle of the projected baseline (North = 0$^\circ$, East = 90$^\circ$).
The squared visibility values and error bars are expressed in percents. The statistical and systematic
(from the calibrator star estimated angular size) error contributions are given separately. \label{table_vis}}\\
\hline \hline
JD$\,- 2.45\ 10^6$ & Stations & FOV (") & $N$ & $B$\,(m) & Az.\,($\deg$) & $V^2 \pm$ stat. $\pm$ syst. (\%) & Calibrators\\
\hline
\endfirsthead
\caption{continued.}\\
\hline \hline
JD$\,- 2.45\ 10^6$ & Stations & FOV (") & $N$ & $B$\,(m) & Az.\,($\deg$) & $V^2 \pm$ stat. $\pm$ syst. (\%) & Calibrators\\
\hline
\endhead
\hline
2216.8643 & UT1-UT3 & 0.15 & 35 & 96.350 & 179.21 & $ 0.50 \pm 0.04 \pm 0.01 $ & $\gamma^{2}$\,Vol \\
2216.8666 & UT1-UT3 & 0.15 & 38 & 96.353 & 179.79 & $ 0.45 \pm 0.07 \pm 0.01 $ & $\gamma^{2}$\,Vol \\
2246.8287 & UT1-UT3 & 0.37 & 108 & 95.906 & 10.59 & $ 1.01 \pm 0.11 \pm 0.03 $ & $\gamma^{2}$\,Vol. $\beta$\,Car \\
2246.8310 & UT1-UT3 & 0.37 & 73 & 95.857 & 11.13 & $ 0.85 \pm 0.18 \pm 0.02 $ & $\gamma^{2}$\,Vol. $\beta$\,Car \\
2302.8796 & E0-G0 & 1.40 & 183 & 14.264 & 102.33 & $ 21.19 \pm 1.98 \pm 0.05 $ & HR\,4630 \\
2302.8851 & E0-G0 & 1.40 & 147 & 14.139 & 104.15 & $ 19.96 \pm 2.27 \pm 0.05 $ & HR\,4630 \\
2304.8225 & UT1-UT3 & 0.12 & 55 & 85.216 & 46.55 & $ 2.36 \pm 0.10 \pm 0.05 $ & HR\,4546 \\
2304.8239 & UT1-UT3 & 0.12 & 203 & 85.033 & 46.87 & $ 2.00 \pm 0.06 \pm 0.04 $ & HR\,4546 \\
2451.5692 & E0-G1 & 1.40 & 83 & 62.115 & 6.82 & $ 2.22 \pm 0.29 \pm 0.07 $ & $\theta$\,Cen \\
2452.5393 & E0-G1 & 1.40 & 29 & 62.229 & 0.10 & $ 2.73 \pm 0.33 \pm 0.07 $ & HR\,4050 \\
2452.5426 & E0-G1 & 1.40 & 142 & 62.227 & 0.90 & $ 2.87 \pm 0.35 \pm 0.08 $ & HR\,4050 \\
2452.5504 & E0-G1 & 1.40 & 253 & 62.209 & 2.84 & $ 2.32 \pm 0.11 \pm 0.06 $ & HR\,4050 \\
2453.5382 & E0-G1 & 1.40 & 30 & 62.228 & 0.49 & $ 2.45 \pm 0.58 \pm 0.07 $ & HR\,4050 \\
2453.5480 & E0-G1 & 1.40 & 224 & 62.207 & 2.93 & $ 2.54 \pm 0.12 \pm 0.08 $ & HR\,4050 \\
2453.5521 & E0-G1 & 1.40 & 113 & 62.190 & 3.93 & $ 2.54 \pm 0.31 \pm 0.08 $ & HR\,4050 \\
2675.8452 & B3-D1 & 1.40 & 139 & 21.815 & 97.97 & $ 12.89 \pm 1.15 \pm 0.04 $ & HR\,4050 \\
2675.8537 & B3-D1 & 1.40 & 221 & 21.540 & 100.74 & $ 13.11 \pm 0.63 \pm 0.04 $ & HR\,4050 \\
2675.8599 & B3-D1 & 1.40 & 111 & 21.331 & 102.80 & $ 12.36 \pm 1.19 \pm 0.04 $ & HR\,4050 \\
2675.9027 & B3-D1 & 1.40 & 237 & 19.763 & 117.77 & $ 12.09 \pm 0.60 \pm 0.03 $ & HR\,4050 \\
2675.9095 & B3-D1 & 1.40 & 108 & 19.501 & 120.31 & $ 12.67 \pm 1.56 \pm 0.03 $ & HR\,4050 \\
2624.7905 & B3-C3 & 1.40 & 466 & 7.932 & 37.64 & $ 26.38 \pm 1.15 \pm 0.01 $ & HR\,4050 \\
2624.7948 & B3-C3 & 1.40 & 373 & 7.940 & 39.05 & $ 24.84 \pm 1.18 \pm 0.01 $ & HR\,4050 \\
2624.7988 & B3-C3 & 1.40 & 390 & 7.947 & 40.37 & $ 24.35 \pm 1.15 \pm 0.01 $ & HR\,4050 \\
2624.8400 & B3-C3 & 1.40 & 144 & 7.986 & 53.48 & $ 25.79 \pm 1.94 \pm 0.01 $ & HR\,4050 \\
2624.8581 & B3-C3 & 1.40 & 356 & 7.978 & 59.09 & $ 25.61 \pm 1.13 \pm 0.01 $ & HR\,4050 \\
2626.7952 & B3-C3 & 1.40 & 333 & 7.950 & 40.99 & $ 22.91 \pm 0.95 \pm 0.01 $ & HR\,4050 \\
2626.8014 & B3-C3 & 1.40 & 337 & 7.959 & 42.97 & $ 23.34 \pm 0.88 \pm 0.01 $ & HR\,4050 \\
2626.8141 & B3-C3 & 1.40 & 349 & 7.974 & 47.06 & $ 23.31 \pm 0.78 \pm 0.01 $ & HR\,4050 \\
2626.8453 & B3-C3 & 1.40 & 468 & 7.983 & 56.82 & $ 25.40 \pm 0.68 \pm 0.01 $ & HR\,4050 \\
2626.8496 & B3-C3 & 1.40 & 421 & 7.981 & 58.15 & $ 26.07 \pm 0.78 \pm 0.01 $ & HR\,4050 \\
2626.8539 & B3-C3 & 1.40 & 381 & 7.977 & 59.49 & $ 25.79 \pm 0.85 \pm 0.01 $ & HR\,4050 \\
2627.8450 & B3-C3 & 1.40 & 235 & 7.982 & 57.58 & $ 24.50 \pm 1.88 \pm 0.01 $ & HR\,4050 \\
2627.8496 & B3-C3 & 1.40 & 283 & 7.978 & 58.99 & $ 25.06 \pm 1.68 \pm 0.01 $ & HR\,4050 \\
2627.8539 & B3-C3 & 1.40 & 313 & 7.974 & 60.32 & $ 25.28 \pm 1.41 \pm 0.01 $ & HR\,4050 \\
2628.8599 & B3-C3 & 1.40 & 277 & 7.960 & 62.97 & $ 24.13 \pm 1.30 \pm 0.01 $ & HR\,4050 \\
2628.8650 & B3-C3 & 1.40 & 147 & 7.950 & 64.53 & $ 26.82 \pm 2.33 \pm 0.01 $ & HR\,4050 \\
2630.8258 & B3-C3 & 1.40 & 139 & 7.986 & 54.17 & $ 24.57 \pm 2.53 \pm 0.01 $ & HR\,4050 \\
2630.8601 & B3-C3 & 1.40 & 372 & 7.949 & 64.71 & $ 23.47 \pm 0.98 \pm 0.01 $ & HR\,4050 \\
2631.8387 & B3-C3 & 1.40 & 205 & 7.978 & 59.00 & $ 24.13 \pm 1.14 \pm 0.01 $ & HR\,4050 \\
2631.8428 & B3-C3 & 1.40 & 90 & 7.974 & 60.25 & $ 24.40 \pm 2.75 \pm 0.01 $ & HR\,4050 \\
2631.8780 & B3-C3 & 1.40 & 151 & 7.888 & 70.97 & $ 24.46 \pm 1.10 \pm 0.01 $ & HR\,4050 \\
2651.8286 & B3-C3 & 1.40 & 435 & 7.867 & 72.53 & $ 27.06 \pm 1.01 \pm 0.01 $ & HR\,4050 \\
2651.8534 & B3-C3 & 1.40 & 195 & 7.740 & 80.02 & $ 25.86 \pm 1.93 \pm 0.01 $ & HR\,4050 \\
2651.8619 & B3-C3 & 1.40 & 421 & 7.684 & 82.62 & $ 25.79 \pm 1.09 \pm 0.01 $ & HR\,4050 \\
2651.8702 & B3-C3 & 1.40 & 332 & 7.625 & 85.15 & $ 24.29 \pm 1.13 \pm 0.01 $ & HR\,4050 \\
2652.8461 & B3-C3 & 1.40 & 185 & 7.767 & 78.64 & $ 28.03 \pm 1.38 \pm 0.01 $ & HR\,4050 \\
2652.8640 & B3-C3 & 1.40 & 142 & 7.650 & 84.11 & $ 26.78 \pm 2.61 \pm 0.01 $ & HR\,4050 \\
2654.7670 & B3-C3 & 1.40 & 88 & 7.984 & 56.25 & $ 24.93 \pm 0.98 \pm 0.01 $ & HR\,4050 \\
2654.7718 & B3-C3 & 1.40 & 272 & 7.982 & 57.73 & $ 23.59 \pm 0.79 \pm 0.01 $ & HR\,4050 \\
2654.7769 & B3-C3 & 1.40 & 120 & 7.977 & 59.31 & $ 28.27 \pm 2.99 \pm 0.01 $ & HR\,4050 \\
2654.8272 & B3-C3 & 1.40 & 71 & 7.837 & 74.59 & $ 26.90 \pm 1.44 \pm 0.01 $ & HR\,4050 \\
2654.8311 & B3-C3 & 1.40 & 335 & 7.818 & 75.77 & $ 24.46 \pm 0.83 \pm 0.01 $ & HR\,4050 \\
2654.8483 & B3-C3 & 1.40 & 435 & 7.720 & 80.99 & $ 26.62 \pm 0.57 \pm 0.01 $ & HR\,4050 \\
2663.7817 & B3-D1 & 1.40 & 212 & 23.795 & 68.34 & $ 15.19 \pm 0.91 \pm 0.05 $ & HR\,4050 \\
2664.8213 & B3-D1 & 1.40 & 66 & 23.197 & 81.14 & $ 12.54 \pm 1.77 \pm 0.04 $ & HR\,4050 \\
2664.8519 & B3-D1 & 1.40 & 187 & 22.493 & 90.57 & $ 14.36 \pm 1.33 \pm 0.05 $ & HR\,4050 \\
2664.8561 & B3-D1 & 1.40 & 187 & 22.379 & 91.88 & $ 14.41 \pm 1.23 \pm 0.05 $ & HR\,4050 \\
2665.8654 & B3-D1 & 1.40 & 227 & 22.032 & 95.70 & $ 12.34 \pm 0.80 \pm 0.04 $ & HR\,4050 \\
2670.6898 & B3-D1 & 1.40 & 320 & 23.946 & 45.81 & $ 13.89 \pm 0.72 \pm 0.05 $ & HR\,4050 \\
2670.7270 & B3-D1 & 1.40 & 263 & 23.985 & 57.49 & $ 16.20 \pm 1.43 \pm 0.06 $ & HR\,4050 \\
2670.7312 & B3-D1 & 1.40 & 287 & 23.976 & 58.77 & $ 14.12 \pm 1.24 \pm 0.05 $ & HR\,4050 \\
2670.7567 & B3-D1 & 1.40 & 217 & 23.845 & 66.57 & $ 16.16 \pm 1.34 \pm 0.06 $ & HR\,4050 \\
2670.7661 & B3-D1 & 1.40 & 177 & 23.762 & 69.40 & $ 15.69 \pm 1.75 \pm 0.05 $ & HR\,4050 \\
2670.8071 & B3-D1 & 1.40 & 172 & 23.154 & 81.81 & $ 13.67 \pm 1.46 \pm 0.04 $ & HR\,4050 \\
2670.8121 & B3-D1 & 1.40 & 158 & 23.051 & 83.35 & $ 14.16 \pm 1.12 \pm 0.05 $ & HR\,4050 \\
2675.8452 & B3-D1 & 1.40 & 139 & 21.815 & 97.97 & $ 12.89 \pm 1.15 \pm 0.04 $ & HR\,4050 \\
2675.8537 & B3-D1 & 1.40 & 221 & 21.540 & 100.74 & $ 13.11 \pm 0.63 \pm 0.04 $ & HR\,4050 \\
2675.8599 & B3-D1 & 1.40 & 111 & 21.331 & 102.80 & $ 12.36 \pm 1.19 \pm 0.04 $ & HR\,4050 \\
2675.9027 & B3-D1 & 1.40 & 237 & 19.763 & 117.77 & $ 12.09 \pm 0.60 \pm 0.03 $ & HR\,4050 \\
2675.9095 & B3-D1 & 1.40 & 108 & 19.501 & 120.31 & $ 12.67 \pm 1.56 \pm 0.03 $ & HR\,4050 \\
2677.7022 & B3-D1 & 1.40 & 458 & 23.993 & 55.70 & $ 14.67 \pm 0.45 \pm 0.05 $ & HR\,4050 \\
2677.7094 & B3-D1 & 1.40 & 466 & 23.982 & 57.94 & $ 14.56 \pm 0.42 \pm 0.05 $ & HR\,4050 \\
2677.7168 & B3-D1 & 1.40 & 463 & 23.961 & 60.21 & $ 13.88 \pm 0.41 \pm 0.05 $ & HR\,4050 \\
2677.7542 & B3-D1 & 1.40 & 389 & 23.684 & 71.59 & $ 12.47 \pm 0.40 \pm 0.04 $ & HR\,4050 \\
2677.7617 & B3-D1 & 1.40 & 357 & 23.590 & 73.85 & $ 12.07 \pm 0.43 \pm 0.04 $ & HR\,4050 \\
2677.7698 & B3-D1 & 1.40 & 377 & 23.473 & 76.31 & $ 12.32 \pm 0.41 \pm 0.04 $ & HR\,4050 \\
2678.8376 & B3-D1 & 1.40 & 485 & 21.795 & 98.17 & $ 13.91 \pm 0.32 \pm 0.04 $ & HR\,4050 \\
2678.8447 & B3-D1 & 1.40 & 450 & 21.565 & 100.49 & $ 13.19 \pm 0.32 \pm 0.04 $ & HR\,4050 \\
2678.8519 & B3-D1 & 1.40 & 386 & 21.326 & 102.85 & $ 12.32 \pm 0.35 \pm 0.04 $ & HR\,4050 \\
2678.8914 & B3-D1 & 1.40 & 389 & 19.881 & 116.62 & $ 14.84 \pm 0.40 \pm 0.04 $ & HR\,4050 \\
2678.8979 & B3-D1 & 1.40 & 208 & 19.634 & 119.02 & $ 15.38 \pm 0.60 \pm 0.04 $ & HR\,4050 \\
2679.8071 & B3-D1 & 1.40 & 387 & 22.594 & 89.36 & $ 13.53 \pm 0.32 \pm 0.04 $ & HR\,4050 \\
2679.8149 & B3-D1 & 1.40 & 256 & 22.386 & 91.81 & $ 12.34 \pm 0.38 \pm 0.04 $ & HR\,4050 \\
2679.8216 & B3-D1 & 1.40 & 285 & 22.197 & 93.92 & $ 12.87 \pm 0.52 \pm 0.04 $ & HR\,4050 \\
2679.8580 & B3-D1 & 1.40 & 355 & 21.018 & 105.82 & $ 13.10 \pm 0.39 \pm 0.04 $ & HR\,4050 \\
2679.8644 & B3-D1 & 1.40 & 113 & 20.789 & 108.00 & $ 13.88 \pm 2.00 \pm 0.04 $ & HR\,4050 \\
2679.8941 & B3-D1 & 1.40 & 193 & 19.674 & 118.63 & $ 15.15 \pm 0.90 \pm 0.04 $ & HR\,4050 \\
2683.7023 & B3-D1 & 1.40 & 206 & 23.954 & 60.78 & $ 11.48 \pm 0.86 \pm 0.03 $ & HR\,4050 \\
2683.7105 & B3-D1 & 1.40 & 128 & 23.916 & 63.31 & $ 13.09 \pm 1.79 \pm 0.03 $ & HR\,4050 \\
2683.7171 & B3-D1 & 1.40 & 169 & 23.875 & 65.31 & $ 12.65 \pm 1.20 \pm 0.03 $ & HR\,4050 \\
2683.7274 & B3-D1 & 1.40 & 430 & 23.792 & 68.45 & $ 12.83 \pm 0.46 \pm 0.03 $ & HR\,4050 \\
2683.7347 & B3-D1 & 1.40 & 408 & 23.719 & 70.63 & $ 12.55 \pm 0.49 \pm 0.03 $ & HR\,4050 \\
2683.7421 & B3-D1 & 1.40 & 437 & 23.631 & 72.89 & $ 13.48 \pm 0.48 \pm 0.03 $ & HR\,4050 \\
2683.7746 & B3-D1 & 1.40 & 431 & 23.094 & 82.72 & $ 12.31 \pm 0.43 \pm 0.03 $ & HR\,4050 \\
2683.7819 & B3-D1 & 1.40 & 467 & 22.936 & 84.96 & $ 13.01 \pm 0.44 \pm 0.03 $ & HR\,4050 \\
2683.7888 & B3-D1 & 1.40 & 444 & 22.775 & 87.10 & $ 12.72 \pm 0.45 \pm 0.03 $ & HR\,4050 \\
2683.8185 & B3-D1 & 1.40 & 243 & 21.963 & 96.43 & $ 12.14 \pm 0.50 \pm 0.03 $ & HR\,4050 \\
2683.8254 & B3-D1 & 1.40 & 105 & 21.749 & 98.64 & $ 13.00 \pm 1.64 \pm 0.03 $ & HR\,4050 \\
2683.8581 & B3-D1 & 1.40 & 192 & 20.621 & 109.59 & $ 12.09 \pm 0.64 \pm 0.03 $ & HR\,4050 \\
2683.8658 & B3-D1 & 1.40 & 449 & 20.334 & 112.31 & $ 13.39 \pm 0.46 \pm 0.03 $ & HR\,4050 \\
2683.8736 & B3-D1 & 1.40 & 256 & 20.040 & 115.10 & $ 12.48 \pm 0.93 \pm 0.03 $ & HR\,4050 \\
2683.8928 & B3-D1 & 1.40 & 172 & 19.306 & 122.25 & $ 14.13 \pm 1.31 \pm 0.04 $ & HR\,4050 \\
2683.9008 & B3-D1 & 1.40 & 272 & 18.998 & 125.39 & $ 13.21 \pm 0.83 \pm 0.03 $ & HR\,4050 \\
2683.9065 & B3-D1 & 1.40 & 275 & 18.784 & 127.63 & $ 13.83 \pm 0.75 \pm 0.04 $ & HR\,4050 \\
2684.7819 & B3-D1 & 1.40 & 211 & 22.874 & 85.81 & $ 13.05 \pm 0.59 \pm 0.01 $ & HR\,4050 \\
2684.7929 & B3-D1 & 1.40 & 80 & 22.607 & 89.21 & $ 11.33 \pm 1.71 \pm 0.01 $ & HR\,4050 \\
2684.7992 & B3-D1 & 1.40 & 118 & 22.442 & 91.16 & $ 12.58 \pm 1.44 \pm 0.01 $ & HR\,4050 \\
2684.8310 & B3-D1 & 1.40 & 198 & 21.476 & 101.38 & $ 12.17 \pm 0.49 \pm 0.01 $ & HR\,4050 \\
2741.7918 & B3-M0 & 1.40 & 55 & 84.580 & 131.47 & $ 1.12 \pm 0.37 \pm 0.04 $ & HR\,4526 \\
2742.7684 & B3-M0 & 1.40 & 70 & 90.319 & 122.20 & $ 0.80 \pm 0.23 \pm 0.03 $ & HR\,4526 \\
2742.7849 & B3-M0 & 1.40 & 52 & 85.714 & 129.50 & $ 2.52 \pm 1.12 \pm 0.09 $ & HR\,4526 \\
2745.6829 & B3-M0 & 1.40 & 36 & 112.435 & 93.90 & $ 0.68 \pm 0.71 \pm 0.01 $ & HR\,4831. HR\,4546 \\
2769.6173 & B3-M0 & 1.40 & 104 & 112.447 & 93.89 & $ 0.81 \pm 0.12 \pm 0.01 $ & HR\,4831 \\
2769.6249 & B3-M0 & 1.40 & 94 & 110.408 & 96.36 & $ 0.65 \pm 0.13 \pm 0.01 $ & HR\,4831 \\
2769.6299 & B3-M0 & 1.40 & 111 & 109.049 & 98.01 & $ 0.82 \pm 0.12 \pm 0.01 $ & HR\,4831 \\
2770.6169 & B3-M0 & 1.40 & 82 & 111.816 & 94.66 & $ 0.85 \pm 0.21 \pm 0.01 $ & HR\,4831 \\
2770.6224 & B3-M0 & 1.40 & 64 & 110.358 & 96.42 & $ 0.94 \pm 0.33 \pm 0.01 $ & HR\,4831 \\
2770.6280 & B3-M0 & 1.40 & 81 & 108.811 & 98.29 & $ 0.63 \pm 0.19 \pm 0.01 $ & HR\,4831 \\
2786.6480 & B3-M0 & 1.40 & 226 & 90.383 & 122.10 & $ 1.47 \pm 0.08 \pm 0.05 $ & HR\,4546 \\
2786.6581 & B3-M0 & 1.40 & 184 & 87.543 & 126.48 & $ 1.50 \pm 0.10 \pm 0.05 $ & HR\,4546 \\
2786.6634 & B3-M0 & 1.40 & 133 & 86.084 & 128.87 & $ 1.49 \pm 0.13 \pm 0.05 $ & HR\,4546 \\
2790.5635 & E0-G0 & 1.40 & 480 & 13.813 & 108.79 & $ 25.20 \pm 0.39 \pm 0.01 $ & HR\,4546 \\
2790.5686 & E0-G0 & 1.40 & 369 & 13.688 & 110.56 & $ 21.89 \pm 0.40 \pm 0.01 $ & HR\,4546 \\
2790.5733 & E0-G0 & 1.40 & 346 & 13.571 & 112.23 & $ 23.13 \pm 0.47 \pm 0.01 $ & HR\,4546 \\
2790.6054 & E0-G0 & 1.40 & 426 & 12.755 & 124.14 & $ 26.06 \pm 0.59 \pm 0.01 $ & HR\,4546 \\
2790.6103 & E0-G0 & 1.40 & 479 & 12.633 & 126.03 & $ 28.03 \pm 0.58 \pm 0.01 $ & HR\,4546 \\
2790.6151 & E0-G0 & 1.40 & 406 & 12.512 & 127.94 & $ 24.87 \pm 0.56 \pm 0.01 $ & HR\,4546 \\
2791.4925 & E0-G0 & 1.40 & 468 & 15.214 & 86.68 & $ 25.02 \pm 0.30 \pm 0.01 $ & HR\,4546 \\
2791.4974 & E0-G0 & 1.40 & 358 & 15.135 & 88.20 & $ 22.31 \pm 0.36 \pm 0.01 $ & HR\,4546 \\
2791.5023 & E0-G0 & 1.40 & 180 & 15.053 & 89.70 & $ 22.05 \pm 0.85 \pm 0.01 $ & HR\,4546 \\
2791.5296 & E0-G0 & 1.40 & 419 & 14.526 & 98.37 & $ 23.30 \pm 0.35 \pm 0.01 $ & HR\,4546 \\
2791.5347 & E0-G0 & 1.40 & 297 & 14.416 & 100.03 & $ 21.75 \pm 0.45 \pm 0.01 $ & HR\,4546 \\
2791.5398 & E0-G0 & 1.40 & 154 & 14.305 & 101.69 & $ 20.42 \pm 0.96 \pm 0.01 $ & HR\,4546 \\
2791.5737 & E0-G0 & 1.40 & 445 & 13.494 & 113.32 & $ 25.73 \pm 0.59 \pm 0.01 $ & HR\,4546 \\
2791.5786 & E0-G0 & 1.40 & 381 & 13.371 & 115.08 & $ 24.49 \pm 0.59 \pm 0.01 $ & HR\,4546 \\
2791.5840 & E0-G0 & 1.40 & 329 & 13.233 & 117.06 & $ 20.99 \pm 0.58 \pm 0.01 $ & HR\,4546 \\
2791.6264 & E0-G0 & 1.40 & 453 & 12.166 & 133.73 & $ 25.88 \pm 0.58 \pm 0.01 $ & HR\,4546 \\
2791.6316 & E0-G0 & 1.40 & 375 & 12.044 & 135.90 & $ 25.05 \pm 0.61 \pm 0.01 $ & HR\,4546 \\
2791.6367 & E0-G0 & 1.40 & 358 & 11.927 & 138.09 & $ 23.81 \pm 0.58 \pm 0.01 $ & HR\,4546 \\
2977.8452 & E0-G0 & 1.40 & 390 & 15.962 & 44.66 & $ 24.45 \pm 0.41 \pm 0.01 $ & HR\,4546 \\
2977.8530 & E0-G0 & 1.40 & 333 & 15.980 & 47.14 & $ 24.03 \pm 0.41 \pm 0.01 $ & HR\,4546 \\
3011.7317 & D0-H0 & 1.40 & 111 & 63.563 & 37.94 & $ 4.72 \pm 0.37 \pm 0.07 $ & HR\,4050 \\
\hline
\end{longtable}
\end{small}

\end{document}